\documentclass[]{spie}  

\usepackage[]{graphicx}
\title{Quantum Noise in Gravitational-wave Interferometers:\\Overview and Recent Developments}

\author{Thomas Corbitt and Nergis Mavalvala\supit{1}
\skiplinehalf Massachusetts Institute of
Technology, 175 Albany Street, Cambridge, MA 02139}

\authorinfo{\supit{1}\noindent Send correspondence to Nergis Mavalvala.
E-mail: nergis@ligo.mit.edu, Telephone: 1 617 253-5657}

\begin{document}
  \maketitle

\begin{abstract}
We present an overview of quantum noise in gravitational wave
interferometers. Gravitational wave detectors are extensively
modified variants of a Michelson interferometer and the quantum
noise couplings are strongly influenced by the interferometer
configuration. We describe recent developments in the treatment of
quantum noise in the complex interferometer configurations of
present-day and future gravitational-wave detectors. In addition,
we explore prospects for the use of squeezed light in future
interferometers, including consideration of the effects of losses,
and the choice of optimal readout schemes.

\end{abstract}



\section{INTRODUCTION}
\label{sect:intro}  

\noindent With several gravitational wave (GW) observatories
worldwide nearing completion \cite{GWifos}, the direct detection
of gravitational waves (GWs) from astrophysical sources is a
rapidly growing enterprise. The Laser Interferometer
Gravitational-wave Observatory (LIGO) is the US component of this
worldwide effort. Even as the first generation detectors are being
commissioned \cite{spie1}, design and planning is well underway
for next-generation detectors, with installation to begin as early
as 2007. The initial detectors have their maximal strain
sensitivity between 40~Hz and 7~kHz, while the advanced detectors
are expected to be a factor of 10 to 15 more sensitive in a band
from 10~Hz to 7~kHz \cite{spie2}. Third generation interferometers
are still in the earliest stages of theoretical
development~\cite{speed1} and table-top testing~\cite{speed2}. In
this paper we will systematically survey the {\it quantum
mechanical} noise in a broad sampling of these interferometers and
discuss the implications for the increasingly important role that
quantum optics and quantum information are likely to play in
advanced GW interferometers.

\noindent Most gravitational-wave (GW) interferometers, such as
those used LIGO, are variants of a Michelson interferometer. The
Michelson interferometer makes it natural to decompose the optical
fields and the corresponding motion of the arm-cavity mirrors into
symmetric and antisymmetric modes. Since the Michleson is operated
on the dark fringe, ideally only optical signals induced by the
antisymmetric motion of the arm-cavity mirrors exit the
antisymmetric -- or output -- port of the beam splitter. The
motion of the mirrors that generates an optical signal at the
output port may be due to a variety of instrumental noise sources
that compete with the mirror displacements a passing GW would
induce.

\noindent Noise sources that impose limits on the strain
sensitivity of the detector include seismic noise due to
terrestrial vibrations at low frequencies, thermal noise due to
thermally driven fluctuations of the mirrors at intermediate
frequencies and shot noise due to fluctuations in photon number at
high frequencies. In the initial detectors these are indeed the
limiting noise sources. In advanced detectors, however, the
sensitivity at almost all frequencies in the detection band is
expected to be limited by {\it quantum} noise, primarily due to
the increase in circulating laser power, but also strongly
influenced by the detailed optical configuration used.

\subsection{Sources of Quantum Noise}
\label{sect:quantnoise}

\noindent Quantum noise refers to a broad class of noise sources
that arise from the quantum nature of the light source and
photodetection process used in GW interferometers. We shall
consider three types of quantum noise in this review:


\begin{itemize}

\item \emph{Radiation-pressure noise} arises from uncertainties
in the mirror positions due to quantum fluctuations exerting
fluctuating radiation pressure on the mirrors. It was shown over
two decades ago that this radiation pressure force can be
attributed to vacuum fluctuations that enter the unused ports of
the interferometer \cite{caves1,caves2}. We shall continue our
analysis in this framework.

\item \emph{Shot noise} arises from uncertainty due to
quantum mechanical fluctuations in the number of photons at the
interferometer output.

\item \emph{Test-mass quantization} noise arises from uncertainty due to intrinsic
quantization of the position and momenta of the mirrors. However,
it has been shown that non-zero contributions to the detector
output commutator from test-mass (mirror) quantization exactly
cancel the the non-zero contributions from the radiation-pressure
and shot noise commutation relations, and thus the test mass
quantization does not limit the measurement process at all
\cite{TMQuant}.

\end{itemize}

\subsection{The Standard Quantum Limit}
\label{sect:sql}

\noindent The standard quantum limit (SQL) is a well-known and
well-used quantity in quantum optics as well as in quantum
measurement theory. While closely related, the SQL has a different
interpretation in these two fields. For gravitational wave
detectors -- where photons are used to measure the positions of
the test-mass mirrors -- the SQL is obtained by exactly balancing
the radiation-pressure induced position fluctuations with
measurement uncertainty due to fluctuations in the number of
photons, or shot noise~\cite{SQL1}. This balance is nothing but a
statement of the Heisenberg uncertainty principle (HUP) for the
position and momentum of the particle: an initial measurement of
the particle's position, imparts an unknown momentum to it via
radiation pressure, which prevents one from predicting the outcome
of a later position measurement, since the momentum and position
do not commute. This is known as quantum back action. For a free
mirror of mass $M$, the power spectral density of the minimum
uncertainty in its position, given by the SQL, is $S_x^{SQL} =
\frac{8 \hbar}{M \Omega^2}$, where $\Omega$ is the angular
frequency angular frequency at which the measurement is made. More
generally, the SQL is the limit on the accuracy with which any
position-sensing device can determine the position of a free
mass~\cite{SQL2}.

\noindent The SQL in interferometers with input power $I_0$, which
can also be stated as $S_x\,S_p \geq \hbar^2/4$, where $S_x
\propto I_0$ is the power spectral density of position
fluctuations due to the uncertainty in the number of photons (shot
noise) and $S_p \propto 1/(I_0 \Omega^2)$ is due to the backaction
(radiation-pressure noise). We notice that the spectrum of the
shot noise is flat, while that of radiation-pressure noise falls
off as $1/\Omega^2$. This causes these two noise sources to
dominate in different frequency bands. A further analogy between
the position-momentum commutation and amplitude-phase commutation
in quantum optics shows that the radiation-pressure noise and shot
noise are associated with two orthogonal quadratures of the
radiation field.

\noindent In initial LIGO interferometers, the input laser power
is low enough that the sensitivity will be limited by shot noise,
and radiation-pressure noise will be completely insignificant.

\subsection{Quantum Non-demolition}
\label{sect:qnd}

A fundamental premise of the SQL is that is that the optical noise
sources -- radiation-pressure noise and photon shot noise --
together enforce the SQL {\it only if they are uncorrelated}. The
SQL can be overcome by creating correlations between the
radiation-pressure and the shot noise. Quantum non-demolition
(QND) devices, first proposed by Braginsky~\cite{qndorig}, are
generically measurement apparatuses that prevent their own quantum
properties from 'demolishing' the state of the system they are
performing a measurement on. QND interferometers are achieved by
creating correlations between the radiation-pressure and
shot-noise.

\noindent Here we discuss three ways in which QND interferometers
can be realized:


\begin{itemize}

\item {\bf Squeezed state injection}\\
Squeezed states of light reduce the noise in one quadrature at the
expense of additional noise in the orthogonal quadrature. As
mentioned above, vacuum fluctuations entering the unused port of
the beamsplitter corrupt the measurement of the mirror position.
Injecting squeezed vacuum with the appropriate squeeze quadrature
into the unused port can reduce the dominant optical noise source,
typically by a factor $S_h \propto {\rm e}^{-2 R}$, where ${\rm
e}^{-2 R}$ is the {\it power squeeze factor}. The {\it squeeze
angle} governs the degree of squeezing in each quadrature.

\item {\bf Ponderomotive squeezing and the variational readout}\\
Ponderomotive squeezing arises from the naturally occurring
correlation of light intensity fluctuations (radiation-pressure
noise) to mirror position fluctuations (shot noise) upon
reflection of light from a mirror. A heuristic description of this
process is that when light (or vacuum) with fluctuations in the
amplitude (radiation-pressure) quadrature, $\Delta I$, and with
fluctuations in phase quadrature, $\Delta \phi$, is incident on a
mirror, the mirror position is influenced by $\Delta I$ due to the
back-action force of the light on the mirror. If the position
signal is measured in the phase quadrature, then the noise on that
measurement is given by $\Delta \phi - \kappa(\Omega) \Delta I$,
where $\kappa(\Omega)$ is measure of the back-action coupling and
{\it depends on the frequency of oscillation of the mirror,
$\Omega$}. If a single quadrature is measured at the output of the
interferometer, the noise on that measurement will depend on
$\kappa(\Omega)$ at each frequency $\Omega$. If, however, one
could measure an admixture of quadratures with a {\it
frequency-dependent} homodyne angle that is a function of
$\kappa$, it is possible to completely eliminate $\Delta I$ from
the measurement, at all frequencies~\cite{KLMTV}.

\item {\bf Optical springs}\\
Another way to to realize QND in an interferometer is by
modification of the interferometer mirror dynamics by coupling to
the light. The basis of this coupling is the fact that the
radiation-pressure force not only imposes random fluctuations on
the positions of the interferometer mirrors, but also exerts a
restoring force with a deterministic frequency-dependent spring
constant (sometimes called 'ponderomotive rigidity')~\cite{BC2}.
The high-power optical field incident on the arm cavity mirrors of
certain interferometer configurations gives rise to mirror
dynamics that can be characterized by a pair of resonances: a
low-frequency resonance due to the {\it mechanical} restoring
force from the ponderomotive rigidity, and a higher-frequency {\it
optical} resonance due to the light storage in the cavity.

\end{itemize}

\section{INTERFEROMETER CONFIGURATIONS AND QND}
\label{sect:ifoconfigs}

\noindent The mechanisms by which the quantum noise couples to the
GW signal at output port of the detector are strongly influenced
by the interferometer configuration and the readout method used.
Since the very earliest analyses of quantum noise in
interferometers, the radiation-pressure noise and shot noise were
assumed to be uncorrelated~\cite{caves1,strain1}. This is largely
true for a Michelson interferometer and certain variations, such
as the power-recycled interferometer used in initial LIGO.
However, with increased input and higher circulating power and the
addition of a "signal recycling" optical cavity at the output port
of the interferometer, it is possible to build up dynamical
correlations between the shot noise and backaction
noise~\cite{BC1}.

\noindent In order to fully explore the quantum noise in
interferometers, detailed knowledge of the interferometer
configuration is therefore necessary. In this section we outline a
few configurations that are particularly useful to study, either
because they are naturally QND interferometers or have the
potential to be converted into QND interferometers.

\begin{figure}[h]
   \begin{center}
   \begin{tabular}{c}
   \includegraphics[height=10cm]{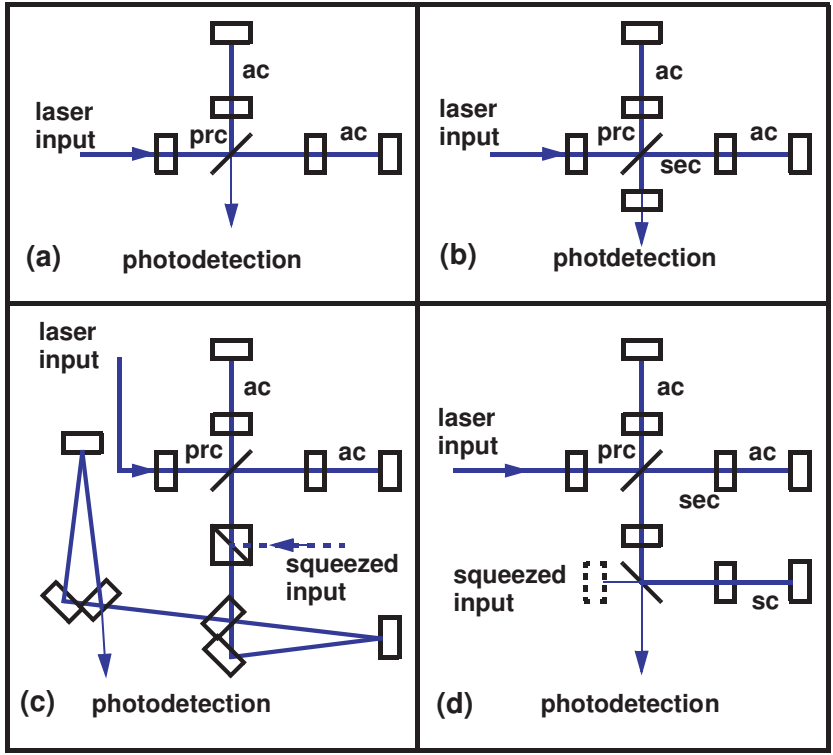}
   \end{tabular}
   \end{center}
   \caption[ifo_configs]
   { \label{fig:ifo_configs}
Schematic representations of some sample interferometer
configurations. In all cases shown except (c) we have chosen
extensions of a power-recycled Michelson interferometer with
Fabry-Perot cavities in each arm as the basic interferometer; this
configuration is shown in (a). (a) Power-recycled Michelson
interferometer with Fabry-Perot arm cavities (PRFPMI); (b)
Signal-tuned PRFPMI; (c) PRFPMI with variational readout and
(optional) squeezed input; and (d) PRFPMI-based speed meter with
(optional) squeezing. Acronyms used: ac = arm cavity, prc =
power-recycling cavity, sec = signal extraction cavity, sc =
sloshing cavity.}
   \end{figure}

\noindent In Fig.~\ref{fig:ifo_configs} we show schematics of some
of the interferometer configurations we refer to in the sections
that follow.

\subsection{Power-recycled interferometers}
\label{sect:prm}

\noindent In the initial LIGO detectors (as well as VIRGO and
TAMA300), two km-scale Fabry-Perot cavities are inserted into the
arms of the Michelson interferometer. Build-up of the optical
field in the cavities enhances the GW-induced phase shift, thus
increasing the sensitivity of the detector. The Michelson
interferometer is operated on the dark fringe to minimize the shot
noise associated with static laser power at the antisymmetric
(dark) port. Since most of the light returns toward the laser, a
partially transmitting power-recycling mirror (PRM) is placed
between the laser source and the beam splitter to `recycle' the
light back into the interferometer~\cite{drever} (see
Fig.~\ref{fig:ifo_configs}(a)).

\noindent The GW signal at frequency $\Omega$, due to the
asymmetric motion of the end mirrors, appears as an phase
modulation at the input coupler (ITM) of the arm cavities. Upon
mixing with a local oscillator field at the beam splitter this
phase modulation is converted to amplitude modulation that is
detected by the photodetector. The GW signal appears {\it only} in
a single quadrature.

\noindent The radiation-pressure noise and shot noise in this
interferometer are uncorrelated, just as in a simple Michelson
interferometer. Vacuum fluctuations entering the antisymmetric
port of the beamsplitter pass into each arm and return to the
antisymmetric port with an (uninteresting) overall phase shift.
Even as the laser power in the interferometer is increased, the
best performance that can be achieved is at the level of the SQL,
unless squeezed light is injected into the unused port of the
beamsplitter. Since there is an absence of QND at all powers under
normal operating conditions, where signal in a single quadrature
is measured, this configuration has has been dubbed "conventional
interferometer". Methods for converting this configuration into a
QND interferometer via by cancellation of radiation-pressure and
shot-noise are discussed in Section~\ref{sect:sqzligoII}.

\subsection{Signal-tuned interferometers}
\label{sect:srifo}

\noindent Signal-tuned interferometers are already used in the
GEO600 detector, and are part of the baseline plan for the
Advanced LIGO detectors.

\noindent The optical configuration currently planned to achieve
quantum-limited performance in Advanced LIGO uses the Resonant
Sideband Extraction (RSE) technique~\cite{RSE}, in addition to
power-recycling. In RSE, an additional partially transmitting
mirror, the signal extraction mirror (SEM), is placed between the
antisymmetric port of the beam splitter and photodetector (see
Fig~\ref{fig:ifo_configs}(b)). The optical properties
(reflectivity, loss) of this signal extraction mirror and its
microscopic position (in fractions of the wavelength of the laser
light, $1.064\,\mu{\rm m}$) can significantly influence the
frequency response of the interferometer~\cite{RSEcomm}. The
resonance condition of the signal extraction cavity -- comprising
the SEM and the input test-mass (ITM) mirrors of the arm cavities
-- and the reflectivity of the SEM control the frequency of peak
response and the bandwidth of the detector, respectively.

\noindent The signal-recycling cavity is offset (detuned) from
resonance to shift the frequency of the peak optical response of
the detector to frequencies where other noise sources are not
dominant. This detuning has the profound consequence, however,
that the frequency response of the detuned configuration is no
longer symmetric around the carrier frequency. As a consequence,
only one of the two (upper or lower) GW-induced sidebands is
exactly on resonance in the signal-recycling cavity. In general,
the the upper and lower GW sidebands contribute asymmetrically to
the total output field, {\it which makes the GW signal appear
simultaneously in {\it both} quadratures of the output
field}~\cite{BC1}.

\begin{figure}[h]
   \begin{center}
   \begin{tabular}{c}
   \includegraphics[height=6.4cm]{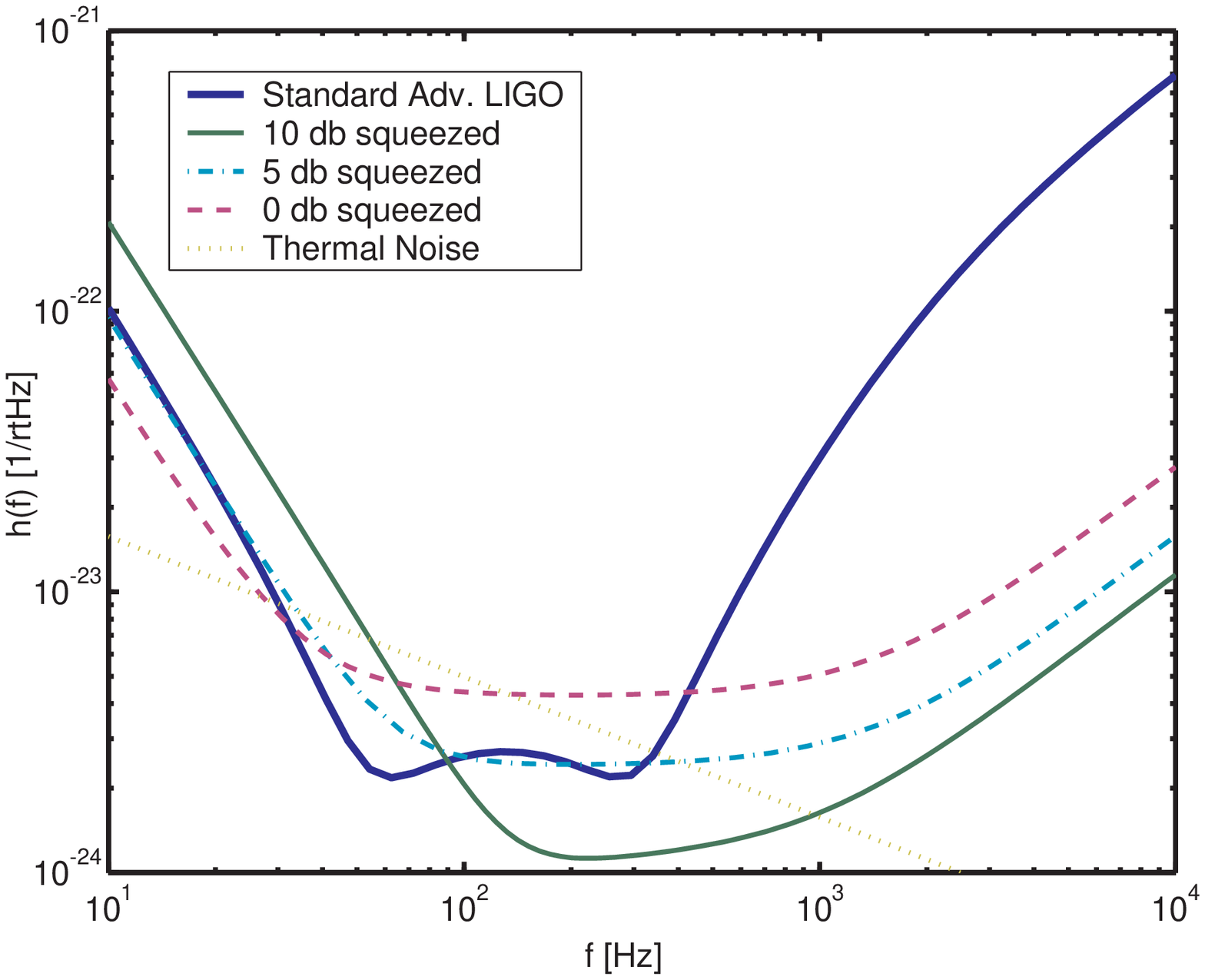}
   \includegraphics[width=7.9cm]{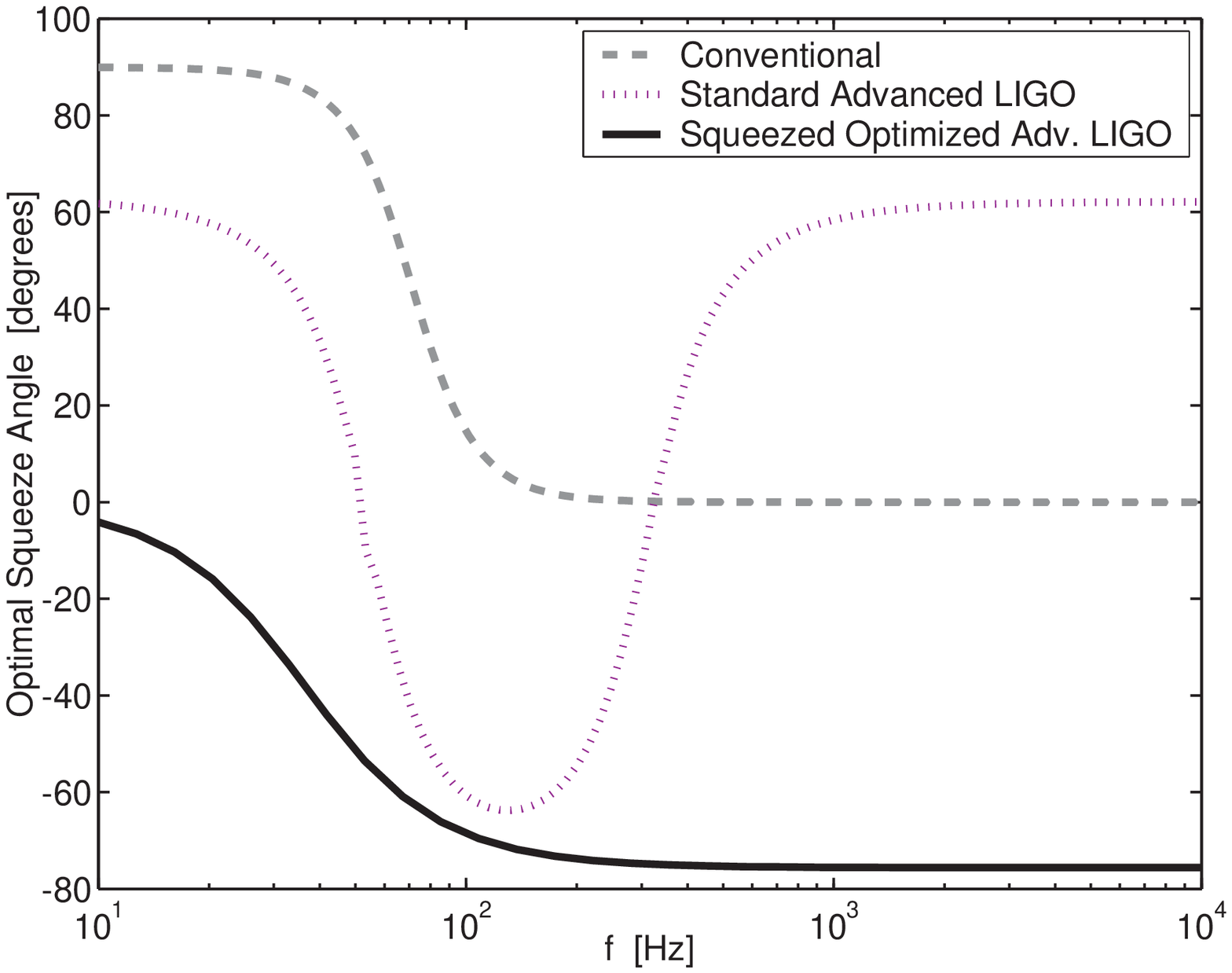}
   \end{tabular}
   \end{center}
   \caption[sqz_adligo_fig01]
   { \label{fig:sqz_adligo}
Right panel: Noise curves for Advanced LIGO with different levels
of squeezing included~\cite{corbitt1}. The dark solid curve is the
standard, or baseline, Advanced LIGO sensitivity when limited by
quantum noise; the dashed (purple) curve corresponds to a
broadband configuration that is better suited for squeezing (but
no squeezing is injected); the dash-dot (cyan) curve is for the
broadband configuration with 5 dB of squeezing; the lighter solid
curve (green) has 10 dB of squeezing injected. The dotted (yellow)
curve is an estimate of the thermal noise, assuming 40 kg silica
test masses limited by internal thermal noise~\cite{spie2}. Left
panel: The optimal squeeze angle for three interferometer
considered. The dashed curve (gray) is for a conventional
power-recycled interferometer; the dotted curve (purple) is for
the standard (baseline) Advanced LIGO configuration; the solid
(black) curve is for the optimized broadband advanced LIGO
configuration.}
   \end{figure}

\noindent To achieve quantum-limited performance, the optical
configuration for Advanced LIGO includes not only the SEM at the
antisymmetric output of the interferometer but also much higher
circulating powers. The higher circulating power immediately
enhances the role of radiation pressure noise in the overall noise
performance of the interferometer. Furthermore, use of the detuned
RSE to optimize the detector response has exposed some surprising
features that are due to the dynamical correlations of the shot
noise and radiation pressure noise~\cite{BC1}. These shot
noise--radiation pressure noise correlations -- which are
manifestations of quantum non-demolition (QND) in that the
correlations lead to below-SQL noise limits -- lead to an
opto-mechanical coupling that significantly modifies the dynamics
of the interferometer mirrors, introducing an additional resonance
at which the sensitivity also peaks (see, e.g., the dark solid
(blue) curve in Fig.~\ref{fig:sqz_adligo})~\cite{BC1,BC2}.

\subsection{Third-generation interferometers}

Several novel interferometer configurations have been proposed for
GW detectors beyond Advanced LIGO. These include all-reflective
interferometers, which circumvent the problems associated with
thermal distortions due to the heat deposited in the absorptive
substrates of mirrors when laser beams pass through them; white
light interferometers which are capable of extremely broadband
operation; and a class of back-action evading interferometers,
known as {\it speed meters}. We will limit our discussion to speed
meters since they are inherently QND devices. Furthermore, we
defer that discussion till Sec.~\ref{sect:sqzligoII}.

\section{Prospects for squeezing}
\label{sect:sqz}

Since the earliest experiments to generate squeezed states of
light were carried out in the 1980's~\cite{slusher,kimble1}, there
has been steady progress in both the degree of squeezing
achieved~\cite{schiller} as well as the stability and long-term
operation of squeezing experiments~\cite{schneider1}. The use of
squeezed light in various applications has nudged squeezed state
generation from delicate laboratory set-ups with millisecond
durations to the realm of stable sources of strongly squeezed
light that operate robustly for several hours at a time. The
present state of the art for CW squeezed light is about 6~dB of
squeezing at several 100~kHz.

\subsection{Squeezed state generation}
For squeezed light to be useful in GW interferometers there are a
few major directions in which the state-of-the-art must progress:


\begin{itemize}

\item {\bf Increased level of squeezing}\\
A benchmark for squeezing to become a viable technology for GW
detectors is squeeze factors of greater than 10 dB. This
requirement is a great challenge but one that has seen steady but
slow progress. One of the most promising techniques for squeeze
source generation continues to be the $\chi^{(2)}$ non-linearity
in crystals. One example of known limitations with $\chi^{(2)}$
non-linear materials is green-induced infrared absorption -- or
GRIIRA -- in LiNbO${\rm _3}$, which is often cited as a limitation
to the degree of squeezing as well as deviations from theoretical
optimizations in OPAs. Absorption measurements of MgO-doped
LiNbO${\rm _3}$ have shown a strong dependence of GRIIRA on both
the stoichiometry and on the MgO concentrations~\cite{GRIIRA}. In
fact, the results indicate that carefully control of these two
effects can all but eliminate GRIIRA (and photorefraction, as an
added side-effect). This is an area of development where new
materials and improvements in the growth parameters of
next-generation crystals will contribute to the elusive 10 dB
goal.

\item {\bf Squeezing at low frequencies}\\
Broadband cw squeezing is presently available at frequencies above
about 300 kHz~\cite{bowen}. This is, of course, not useful for GW
interferometers that need quantum-limited operation at around 100
Hz. There are two primary factors which give rise to the present
frequency limit: (i) Classical noise on the seed beam, i.e. the
seed beam is not shot-noise-limited at frequencies below about 1
MHz; and (ii) Noise introduced inside the non-linear devices used
for squeezing, e.g. in the optical parametric amplifiers (OPAs).

\noindent The first limitation is not particularly relevant for
squeezed {\it vacuum}, which is all that is required for injection
into the unused ports of gravitational-wave interferometers.
Recently, there has also been some progress on the second
limitation, regarding noise generated in OPAs. Classical laser
noise is still a problem in that it couples to the squeezed output
beam via experimental imperfections such as crystal
non-uniformities and mode-mismatch. One approach is to facilitate
noise cancellations by passing both the pump beam and the seed
beam through a pair of OPAs~\cite{bowen}. Careful control of the
relative phases of the beams allows for cancellation of the
classical noise on the laser. This technique has led to realizing
squeezing in the 200 kHz band, but continues to be limited by
technical noise due to the imbalance between the two OPAs and
uncorrelated acoustic noise in the OPAs. A technique for
eliminating this latter is effect is proposed by use of a ring OPA
to classically correlate the noise on each of the squeezed outputs
that are counter-propagating in the ring OPA~\cite{bowen}.

\item {\bf Lowering losses}\\
Another area of development for increasing the squeezing
efficiency is the minimization of other losses in the system.
These include photodetection efficiency and optical losses that
destroy the hard-earned quantum correlations on the squeezed
light. In Section~\ref{sect:sqzligoII}, we highlight the effects
of a variety of losses in a GW detector with squeezing.

\end{itemize}

\subsection{Applications of squeezing to gravitational-wave
interferometers} \label{sect:sqzligoII}

In this section we describe some of the effects of using squeezed
light in GW interferometers. We assume availability of 10 dB of
vacuum squeezing available from DC to 10~kHz. 10 dB corresponds to
the squeeze factor ${\rm e}^{-R}= 0.3$. Here we do not include
losses, but the effect of losses are described in
Sec.~\ref{sect:sqz_losses}. We consider the use of squeezed light
in:


\subsubsection{Power-recycled interferometers}

\begin{figure}[h]
   \begin{center}
   \begin{tabular}{c}
   \includegraphics[height=7cm]{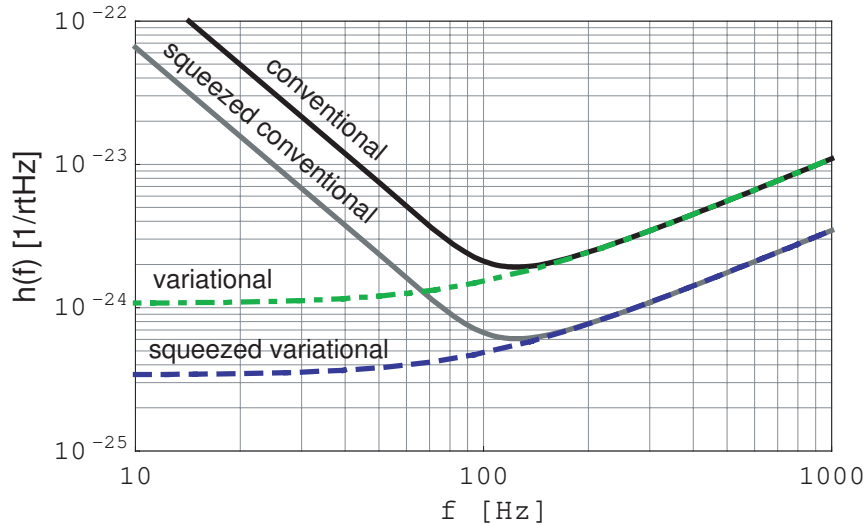}
   \end{tabular}
   \end{center}
   \caption[sqz_adligo_fig01]
   { \label{fig:klmtv_fig01}
Noise curves for a power-recycled interferometer (with initial
LIGO parameters) with squeezing and with a variational
readout~\cite{KLMTV}. The solid dark (black) curve uses the
standard, or baseline, initial LIGO parameters but power at the
beamsplitter that gives SQL-limited performance at 100 Hz, i.e.
($I_0 = I_{SQL} \simeq 10~{\rm kW}$; the solid light (gray) curve
is the sensitivity for the same interferometer but with squeezing
injected at a {\it frequency-dependent} optimal squeeze angle; the
dash-dot (green) curve is for the same interferometer but using a
{\it frequency-dependent} homodyne readout that measures the
optimal quadrature at each frequency; the lighter dashed curve
(blue) uses a variational readout as well as 10 dB of squeezing
injected. No losses are included in the noise curves plotted here,
but the treatment including losses can be found in
Ref.~[\citenum{KLMTV}].}
   \end{figure}

Several variations have been proposed to turn these "conventional"
interferometers into QND devices. The sensitivity curves for these
are shown in Fig.~\ref{fig:klmtv_fig01}.

\begin{itemize}

\item In the \emph{squeezed-input interferometer} squeezed vacuum is
injected into the antisymmetric port of the interferometer.
Squeezing has the effect of increasing the fluctuations in one
quadrature, while decreasing fluctuations in the other. The
\emph{squeeze angle} describes the linear combination of input
quadratures in which the fluctuations are reduced. Since
radiation-pressure noise and shot noise dominate in different
frequency regions and depend on different input quadratures, the
squeeze angle that achieves optimal performance is frequency
dependent. A possible configuration is choosing a frequency
independent squeeze angle that decreases shot noise and increases
radiation pressure noise. This configuration is equivalent to an
unsqueezed interferometer with an increase to the laser power by
the squeeze factor\cite{caves2}. To obtain a broadband performance
increase, two km-scale(required for low losses) Fabry-Perot filter
cavities can produce the required frequency dependent squeeze
angle from a fixed squeeze angle source. In this configuration,
the noise(power) is reduced by the squeeze factor at all
frequencies.

\item In the {\emph squeezed-variational interferometer} two
km-scale Fabry-Perot filter cavities to perform a frequency
dependent homodyne detection of the signal. This detection method
eliminates radiation-pressure noise from the signal by creating
correlations between the shot noise and radiation-pressure noise.
As a result, squeezing must only reduce the shot noise, and the
radiation-pressure noise may be disregarded, which results in a
frequency independent squeeze angle and the noise (power) is
reduced by the squeeze factor at all frequencies.

\end{itemize}

\subsubsection{Signal tuned interferometers}
The results shown in this section are a preview of work that
appears in a manuscript we are preparing for
submission~\cite{corbitt1}. We limit our discussion to the detuned
RSE interferometers planned for Advanced LIGO and show that there
are reasonable gains to be made even with frequency-independent
squeezing.

\noindent The introduction of signal recycling has the effect of
mixing the quadratures in the input/output relations of the
interferometer. This allows the shot noise and radiation pressure
noise to become correlated, producing the two well known
resonances typical of signal recycled interferometers\cite{BC1}.

\noindent To improve the performance of this configuration, we
inject squeezed vacuum into the dark port. The mixing of the
quadratures results in the optimal squeeze angle being modified
from the power-recycled case. It has been shown that this
frequency-dependent squeeze angle can be produced by kilometer
scale filter cavities\cite{KLMTV,GEOSQ}. Due to the inherent
difficulty of using long filter cavities, we will consider the
case of a fixed squeeze angle in the Advanced LIGO configuration.

\noindent For the standard detuning of the signal recycling cavity
in Advanced LIGO, the resonances are placed near 100~Hz to achieve
the best performance in the frequency range with the best
sensitivity. This results in the optimal squeeze angle being
strongly frequency dependent and in making broadband improvements
impossible. By choosing the optimal squeeze angle for a particular
frequency band, narrowband improvements can be achieved, at the
cost of worse performance at other frequencies. To increase the
broadband performance, we must modify the detuning of the signal
recycling cavity so as to reduce the variance of the optimal
squeeze angle. We choose a detuned signal extraction cavity, such
that carrier light with angular frequency $\omega_{l}$ obtains a
net phase shift of $\frac{\pi}{2}$ in one pass through the cavity.
In this configuration, the optimal squeeze angle varies much less
than in the standard configuration. We choose a squeeze angle to
optimize the noise performance at 200~Hz. This choice allows us to
improve the performance over the bulk of our frequency range. The
performance of the squeezed configuration is comparable to the
unsqueezed configuration in the frequency range 10~Hz to 300~Hz,
and dramatically better at 300~Hz to 10~kHz.

\subsubsection{Speed meters}
\noindent The principle of a "speed" meter is most easily
understood in terms of the HUP applied to position and momentum.
In Section~\ref{sect:sql}, we highlighted how an position
measurement at time $t_0$ disturbs a consequent measurement at a
later time $t_1 > t_0$ due to the momentum "kick" from the first
measurement. This back-action obviously does not make position a
very good QND observable. While a measurement of momentum
certainly perturbs the position of a test particle as required by
the HUP, that position "kick" does not influence the time
evolution of the momentum, and hence there is no back-action.
Momentum measurements are thus inherently back-action
evading~\footnote{Momentum, or velocity, is a constant of the
motion for a free particle; it commutes with itself at different
times and is, therefore, a good QND variable.} , provided no
position information is collected. Interferometric measurements
that measure the {\it speed} -- similar in behavior to the {\it
momentum} -- of the mirrors were first proposed by Braginsky et
al.~\cite{speed0} and refined into more practical designs by
Purdue and Chen~\cite{speed1,speed2}.

\noindent Many Michelson-based variants of interferometric speed
meters rely on the addition of a "sloshing" cavity at the output
port of an otherwise position sensitive interferometer. The role
of the "sloshing" cavity is to completely cancel the momentum
"kick" due to the position measurement at $t_0$ by an equal and
opposite "kick" at $t_0 + \tau_s$, where $\tau_s$ is the storage
time of the sloshing cavity. Optically, this occurs due to a $\pi$
phase shift in the coupling constant that connects the sloshing
cavity field to the interferometer field. This behavior is quite
analogous to that of two weakly coupled oscillators: when
something other than a normal mode of the system is excited,
energy "sloshes" between the two oscillators, with a $\pi$ phase
shift after each "slosh" cycle~\cite{PCprivate}. A schematic
representation of a sample speed meter interferometer is shown in
Fig.~\ref{fig:ifo_configs}(d). The unused port of the readout
mirror that connects the sloshing cavity (sc) to the signal
extraction cavity (sec) can with be plugged (with the dashed
mirror), or it can be used for squeezed vacuum injection. We do
not include noise curves corresponding to various speed meter
configurations here, but these appear aplenty in
Ref.~[\citenum{speed1}].


\section{The role of losses}
\label{sect:sqz_losses}

\noindent Optical losses play an important role in the performance
of a squeezed interferometer. There are two mechanisms that must
be considered in simultaneously when evaluating the effects of
losses: (i) The dissipation leads to smaller signals, which is
purely classical effect; and (ii) A lossy port allows ordinary
(unsqueezed) vacuum to enter and superpose on the squeezed field
in the interferometer, thus destroying the effects of the
squeezing.

\noindent In general, the significance of the different losses in
an interferometer or associated injection/readout scheme depends
on the interferometer configuration. Squeezed light injection
losses are obviously problematic, since they correspond to pure
degradation of the level of squeezing. The sensitivity to optical
losses in an interferometer depends on the build-up of signal and
noise fields in the various part of the interferometer. In the
variation and/or squeezed-variational interferometer of
Fig.~\ref{fig:ifo_configs}(c), for example, the losses in the arm
cavities dominate the overall performance of the
detector~\cite{KLMTV}.

\noindent In this section we consider the effect of losses in the
signal-tuned interferometer described in
Sec.~\ref{sect:sqzligoII}, as an example~\cite{corbitt1}. The
effects of the losses is strongly dependent on the buildup of the
noise fields in the signal extraction cavity and in the arm
cavities. There are four types of losses we consider:

\begin{itemize}

\item \emph{Injection losses} are losses associated with the injection
of squeezed light into the dark port. This effectively limits the
squeezing magnitude, and we assume that these are included in the
squeezing magnitudes used and are subsequently ignored.

\item \emph{Detection losses} are due to quantum inefficiencies in the
detection of the signal light. These losses allow a small amount
of vacuum fluctuations to leak into the measurement, decreasing
its precision. The vacuum fluctuations are small relative to the
radiation pressure noise at low frequencies and thus has little
effect where radiation pressure noise dominates. The vacuum
fluctuations are significant relative to show noise, however, and
hurt performance at high frequencies where shot noise dominates.

\item \emph{Signal recycling losses} are produced by the beamsplitter, signal
recycling mirror, and the anti-reflective coatings on the initial
test masses. These losses are either amplified or suppressed
depending on the buildup of the noise fields in the signal
extraction cavity. The frequency-dependent phase shift experienced
by the noise fields in interacting with the arm cavities results
in the signal recycling cavity being resonant at high frequencies,
and anti-resonant at low frequencies, and thus the effects of the
losses is amplified at high frequencies and suppressed at low
frequencies.

\item \emph{Arm cavity losses} are incurred in the arm cavities due to
diffraction and absorption. Due to the suppression of the noise
fields at low frequencies in the signal recycling cavity, and to
the suppression of the noise fields at high frequencies in the arm
cavities, the noise fields do not resonate strongly in the arm
cavities. This effect reduces the importance of arm cavity losses.

\end{itemize}

   \begin{figure}[h]
   \begin{center}
   \begin{tabular}{c}
   \includegraphics[height=7cm]{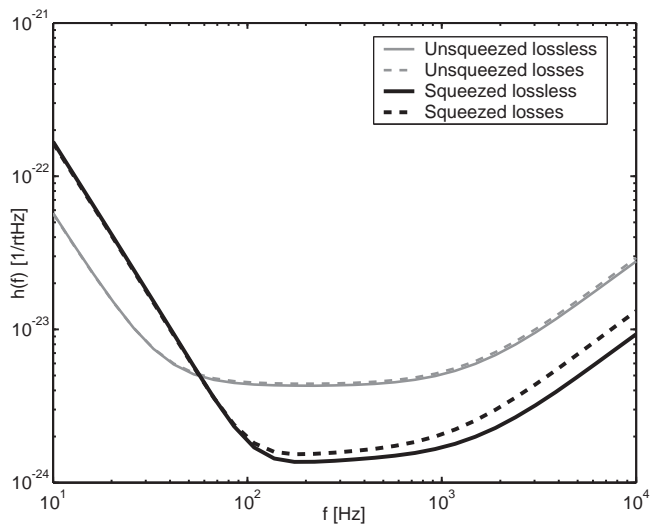}
   \end{tabular}
   \end{center}
   \caption[sqz_adligo_losses]
   { \label{fig:sqz_adligo_losses}
Noise curves for a squeezed input signal-tuned interferometer
(with broadband Advanced LIGO parameters) including
losses~\cite{corbitt1}. The solid gray curve is for the broadband
Advanced LIGO configuration shown as the dashed (purple) curve in
Fig.~\ref{fig:sqz_adligo}; the  dashed gray curve is the
sensitivity for the same interferometer but with additional losses
(see below) ; the solid black curve is for the lossless
interferometer squeezing injected (same as the lighter solid
(green) curve of Fig.~\ref{fig:sqz_adligo}; the dashed curve black
curve includes losses. The losses assumed in these curves are 50
ppm per high-reflection surface, 200 ppm per anti-reflection
surface and photodetection efficiency of 98\%.}
   \end{figure}

\noindent While not apparent from
Fig.~\ref{fig:sqz_adligo_losses}, we have determine that the
dominant losses in this configuration arise from the {\it signal
recycling cavity}, and that the effects of the losses are the
worst at frequencies above 300~Hz and limit the amount of
squeezing that is beneficial at these frequencies. The losses are
largely unimportant at frequencies below 300~Hz. As mentioned
above, these effect of these losses is amplified or suppressed,
depending on the resonance condition in the signal extraction
cavity.

\section{The choice of readout scheme}

\noindent Since GW interferometers operate on a dark fringe, the
intensity of the light exiting the antisymmetric port is quadratic
in the GW amplitude, and therefore insensitive to it, to first
order. The standard way to circumvent this is to interfere the
signal field with a relatively strong local oscillator (LO) field,
such that the intensity of the total optical field, detected at
the beat frequency, varies linearly with the GW amplitude. The
various methods of measuring the GW-induced signal at the
antisymmetric port are referred to as {\it readout schemes.}

\noindent Our discussion thus far has been limited to homodyne
detection, where the beat between the optical signal field at
frequencies $(\omega_l \pm \Omega)$ and the carrier field at
frequency $\omega_l$ is detected (i.e. the LO field oscillates at
exactly the same frequency as the incident laser). Here $\omega_l
\sim 2 \times 10^{15}$ refers to the laser frequency, while
$\Omega \sim 10~{\rm -}~10000~{\rm Hz}$ is the frequency of the
GW. It is worth commenting on the viability of heterodyne readout
schemes, in which the LO has different frequencies from the
carrier. The heterodyne readout is usually implemented, as in
initial LIGO, by using phase modulated light: the light incident
on the interferometer consists of a carrier at $\omega_l$ and
radio frequency (RF) phase modulation (PM) sidebands at
frequencies $\omega_l \pm omega_{RF}$. The PM sidebands are
transmitted with maximum efficiency to the antisymmetric port via
Schnupp asymmetry~\cite{Schnupp}, while the carrier still returns
to the symmetric port. The transmitted sidebands then act as a LO
against which the GW signal can beat, generating signals at
$\pm(\omega_{RF} \pm \Omega)$. Demodulation at $\omega_{RF}$
converts the signal back down into the baseband.

\noindent The heterodyne readout scheme has some advantages: (i)
laser technical noise can be circumvented by upconverting the
signal detection to frequencies where the laser light is
shot-noise-limited (a few MHz); and (ii) more than one quadrature
of the interferometer output can be measured~\footnote{We refer to
this as "variable-quadrature heterodyne detection" after
Ref.~[\citenum{BCM}]}. The latter attribute is of great importance
in detuned RSE interferometers, where the GW signal appears in
{\it both} quadratures, and {\it the optimal detection quadrature
is frequency-dependent}.

\noindent Unfortunately, there is also a fundamental disadvantage:
an \emph{additional} contribution to the quantum noise due to
vacuum fluctuations in frequency bands that are twice the
modulation frequency away from the carrier, as compared with the
homodyne readout scheme. This effect was studied extensively for
detection of a single (phase)
quadrature~\cite{GBL,Schnupp,niebauer,strain1,chickarmane}, which
is adequate for interferometers with low circulating power, and
hence negligible back action noise.

\noindent Recently, the heterodyne readout scheme was generalized
to detection of generic quadratures with LO light that is an
arbitrary mixture of phase and amplitude modulation, more
applicable to advanced GW detectors~\cite{BCM}. This analysis led
to some startling general results. The additional heterodyne noise
is a direct consequence of the HUP: as long as more than one
quadrature is available for simultaneous measurement, the
additional heterodyne noise will appear. Moreover, the HUP imposes
a quantum limit to the additional heterodyne noise, which is
independent of frequency (unless frequency-dependent squeezing is
implemented). This frequency-independent quantum limit puts hard
constraints on the ability of the variable-quadrature optimization
afforded by heterodyne schemes to achieve QND performance over a
broad range of frequencies.

\section{conclusion}

We have attempted to give a very broad survey of recent
developments in our understanding of quantum noise in
gravitational-wave interferometers. With regard to laser
interferometer gravitational-wave detectors, the stage was set in
the late 1970s and early 1980s by the work of Caves, Thorne,
Braginsky et al., and many, many others, which introduced the
conceptual and mathematical formulation of the problem of quantum
noise limits in macroscopic measurements and the possibility of
circumventing them by quantum non-demolition techniques, using
squeezed light, for example. These ideas were deemed not to be
realizable in GW detectors at the time~\footnote{In the
Conclusions section of his 1981 paper~\cite{caves2}, Caves wrote:
"The squeeze-state technique outlined in this paper will not be
easy to implement... Difficult or not [it] might turn out at some
stage to be the only way to improve sensitivity of detectors
designed to detect gravitational waves..."}. In the following
decade, there were significant enough advances in the generation
of non-classical states of
light~\cite{slusher,kimble1,schneider1,lam} and their use to make
modest gains in interferometric measurements below the quantum
limit~\cite{kimble2,mckenzie}, that once again renewed interest in
the topic at the turn of the millennium. The work of Buonanno and
Chen~\cite{BC1,BC2}, building on that of Kimble et
al.~\cite{KLMTV}, led to better understanding of the important
role that naturally occurring quantum correlations can play in
advanced GW detectors with higher power, hence non-negligible back
action noise. At the present time, we have entered an era of
vigorous activity and interest in possibility of sub-SQL
measurement techniques that aim to take advantage of (i)
squeezed-state generation and injection; (ii) the naturally
occurring ponderomotive squeezing in interferometers; (iii) other
back action evading measurement techniques based on speed
meters~\cite{speed1,speed2}; and a variety of other techniques
that we have not described here~\footnote{The authors apologize in
advance to groups and individuals who made pivotal contributions
to the developments mentioned here that we did not have time/scope
to mention.}.

\section*{ACKNOWLEDGMENTS} We thank Stan Whitcomb and Yanbei
Chen for their invaluable comments/corrections to this manuscript,
and our colleagues at the LIGO Laboratory for many valuable
discussions. The LIGO Observatories were constructed by the
California Institute of Technology and Massachusetts Institute of
Technology with funding from the National Science Foundation under
cooperative agreement PHY-9210038. The LIGO Laboratory operates
under cooperative agreement PHY-0107417.


\end{document}